# Large Upper Critical Field and Irreversibility Field in MgB$_2$ wires with SiC additions

M.D. Sumption[1], M. Bhatia[1], E.W. Collings[1], M. Rindfleisch[2],

M. Tomsic[2], S. Soltanian[3] and S.X. Dou[3]

[1] LASM, Materials Science and Engineering Department,

The Ohio State University, Columbus, OH 43210, USA

[2] Hyper Tech Research, Inc. Columbus, OH 43210, USA

[3] ISEM, The University of Wollongong, Wollongong, NSW, Australia

**Abstract**

Resistive transition measurements are reported for MgB$_2$ strands with SiC dopants. The starting Mg powders were 325 mesh 99.9% pure, and the B powders were amorphous, 99.9% pure, and at a typical size of 1–2 µm. The SiC was added as 10 mol% of SiC to 90 mol% of binary MgB$_2$ [(MgB2)0.9(SiC)0.1]. Three different SiC powders were used; the average particle sizes were 200 nm, 30 nm, and 15 nm. The strands were heat treated for times ranging from 5 to 30 minutes at temperatures from 675°C to 1000°C. Strands with 200 nm size SiC additions had $\mu_0 H_{irr}$ and $B_{c2}$ which maximized at 25.4 T and 29.7 T after heating at 800°C for 30 minutes. The highest values were seen for a strand with 15 nm SiC heated at 725°C for 30 minutes which had a $\mu_0 H_{irr}$ of 29 T and a $B_{c2}$ higher than 33 T.

**Introduction**

It has been demonstrated that in some cases the irreversibility field, $\mu_0H_{irr}$, and upper critical field, $B_{c2}$, of $MgB_2$ can be enhanced from the values seem from the binary compound. This has been most evident in thin film results[1,2], with $B_{c2}$ reaching 49 T and even higher at 4.2 K. This is understood to be due to increased scattering in the conductor, an effect seen in a number of materials but more pronounced for $MgB_2$ because of its two-gap nature[3]. A full understanding of just what material modifications enable the $B_{c2}$ enhancement in champion $B_{c2}$ films is lacking, however, one possibility seems to be lattice distortion induced in part by C substitution in the B-sublattice[2]. Several efforts to generate similar $B_{c2}$ enhancements in $MgB_2$ wires and bulks has been stimulated by these results. A number of researchers have investigated C doping in bulks[4,5] to increase $\mu_0H_{irr}$ and $B_{c2}$, which are, of course, strongly correlated (in this context, see[2]), and more generally high field properties in wires with SiC[6-10]. Bulk samples which had been exposed to excess Mg vapor showed high $B_{c2}$ values as well, with a $B_{c2}(0)$ estimate of 29 T[2]. Numerous other additives have also been investigated[11,12], in some cases with the express purpose of changing the B or Mg sublattice, or otherwise changing the electronic state of the system.

Given the variation in preparation conditions and the sensitivity of $MgB_2$ critical fields to $\kappa$, there is some variation in the reported values of $\mu_0H_{irr}$ and $B_{c2}$. Irreversibility fields for *ex-situ* $MgB_2$ tapes have been seen at $\approx$ 12 T for field perpendicular to the tape face[13] (ex-situ tapes are anisotropic such that $H_{//} \approx 1.4\ H_{\perp}$ [14] due to partial alignment of the Mg and B planes parallel to the broad face of the tapes during rolling). Suo and Flukiger found $\mu_0H_{irr}$ values of 8 T and 10.4 T in perpendicular and parallel orientations of the applied field, and, under the same orientations, $B_{c2}$s of 11.9 T and 15.1 T [15,16]. Recent results from Goldacker et al would extrapolate to higher values at 4.2 K[17]. Matsumoto and Kumakura[18] used $SiO_2$ and SiC in the *in-situ* process, enhancing $\mu_0H_{irr}$ from ≈ 17 T to ≈ 23 T at 4 K. $ZrSi_2$, $ZrB_2$ and $WSi_2$ additions were also attempted[19], with some increases seen. Hydride based $MgB_2$ powder with SiC dopants was also investigated, and seen to give $B_{c2}$ values of ≈ 23 T[20]. Dou et al. showed improved high field critical current results for SiC [6-9], and similar wires measured in this laboratory showed improvements in the apparent irreversibility field as might be extrapolated from transport results[10]. In this work, strands of similar construction were investigated systematically, and high field resistive transitions were used to demonstrate relatively large values of $\mu_0H_{irr}$ and $B_{c2}$ with various kinds of SiC dopants under various heating conditions.

**Experimental Methods**

Round, monofilamentary, PIT $MgB_2$ strands were fabricated by continuous (described previously in[10]) and standard powder in tube methods. The outer sheath was Fe enclosed in Cu-30wt%Ni. The starting Mg powders were 325 mesh 99.9% pure, and the B powders were amorphous, 99.9% pure, and at a typical size of 1–2 μm. For strands A and B the powders were V-mixed and then planetary milled[10]. Coarse SiC (200 nm) powders were added during this process in the proportion 10 mol% of SiC to 90 mol% of binary $MgB_2$ [(MgB2)0.9(SiC)0.1].

Strands C and D, fabricated at the University of Wollongong (UoW), were also made from an in-situ route, in this case with 10wt% of "fine" SiC powder (15 and 30 nm). All strands were heated in flowing Ar at temperatures ranging from 640°C to 725°C for 30 minutes. Ramp-up and ramp-down times were short.

Four-point transport measurements were made on 1 cm long samples at the National High Magnetic Field Laboratory in Tallahassee. Standard Pb–Sn solder was used for forming the contacts on the outer sheath, and the distance between the voltage taps was 5 mm. The applied current was 10 mA, and current reversal was used. All measurements were made at 4.2 K in applied fields ranging from 0 to 33 T. The samples were placed perpendicular to the applied field, values of $\mu_0 H_{irr}$ and $B_{c2}$ being obtained taking the 10% and 90% points of the resistive transition. Resistive transitions in self field were used to obtain $T_c$ curves. In this case the samples were from neighboring sections of strand, and were 6 cm long, with voltage taps 3 cm apart.

**Data and Analysis**

Figure 1 shows the resistive transitions for strand A after heating for various times at 700°C and 800°C. It can be clearly seen that both $\mu_0 H_{irr}$ and $B_{c2}$ increase with increasing heating time and temperature in the ranges investigated. The results for strand A are given in Table 1. These results are significantly higher than those of earlier reports for binary $MgB_2$ (for example 8-10.4 T for $\mu_0 H_{irr}$ and 12-15 for $B_{c2}$ [15,16]). In particular, the $\mu_0 H_{irr}$ and $B_{c2}$ values for 800°C and 30 minutes of 25.4 T and 29.7 T represent, to our knowledge, some of the highest $\mu_0 H_{irr}$ and $B_{c2}$ reported for SiC doped $MgB_2$ strands. The $\mu_0 H_{irr}$ values are also higher than might have been expected from the extrapolation of high field critical current results using a Kramer method[10].

The response of strand B is shown in Figure 2, where trends similar to those of strand A are seen. Curves of $\mu_0H_{irr}$ and $B_{c2}$ for strand B heated at various temperatures are plotted vs heating time in Figure 3. In this case a 900°C curve is also present, which is lower than the 800°C curve. Heating at 800°C for 30 minutes gives the highest values, 29.4 T and 31.3 T for $\mu_0H_{irr}$ and $B_{c2}$, respectively. Figure 4 shows the $T_c$ curves (resistive transitions under self field for the B-series samples heated for various times at 800°C. $T_c$ midpoints of 34.2 K, 34.4, 37.8 and 34.4 were found for heating times of 5, 10, 20, and 30 minutes respectively, with transition widths (as measured from 10% to 90% of the transition) of 1.2 to 1.4 K. The $T_c$ value of about 33.2 K (for 10 and 30 minute heatings) corresponds to an expected enhancement of $B_{c2}^2$. However, the fact that the overall resistivity of the strand is not changing drastically, along with the fact that the strand heated for 20 minutes has a $T_c$ of 36.2 indicates significant inhomogeniety in the wire. It is likely that various current paths exist, some of which have different compositions, and no doubt various orientations are being probed as well. This is consistent with the lower $\mu_0H_{irr}$ that has been extrapolated from higher current transport measurements (note that significant "tails" were present)[10].

Figure 5 shows the resistive transitions for strand C and D after various heating temperatures and times. These two strands had the smallest SiC powder sizes (at 15 and 30 nm average size). No clear distinction between the results of these two strands is seen, although as a group they have higher values of $\mu_0H_{irr}$ and $B_{c2}$ than do strands A and B containing the coarse SiC. The highest values are seen for higher temperatures within this set, consistent with the results of strands A and B. The highest values seen were for strand C (15 nm SiC) which had a $\mu_0H_{irr}$ of 29 T and a $B_{c2} > 33$ T.

**Conclusions**

In this work we have presented $\mu_0H_{irr}$ and $B_{c2}$ for $MgB_2$ strands with SiC additions made using powder-in-tube, in-situ powder methods. Higher values of $\mu_0H_{irr}$ and $B_{c2}$ were seen for strands heated at higher temperatures and in some cases longer times. Strands with finer SiC powders also had larger $\mu_0H_{irr}$ and $B_{c2}$ values. In particular, strands with 200 nm size SiC additions had $\mu_0H_{irr}$ and $B_{c2}$ which maximized at 25.4 T and 29.7 T for strands heated at 800°C for 30 minutes. Strands with 15 nm and 30 nm additions had even higher values. The highest critical-field values were seen for a strand with 15 nm SiC additions which after 725°C for 30 minutes had a $\mu_0H_{irr}$ of 29 T and a $B_{c2}$ greater than 33 T.

**Acknowledgements**

This work was supported by a State of Ohio Technology Action Fund Grant and by the US Department of Energy, HEP, Grant No. DE-FGG02-95ER40900.

## List of Tables

Table 1. $B_{c2}$ and $\mu_0 H_{irr}$ for strand A (200 nm SiC).

Table 2. $B_{c2}$ and $\mu_0 H_{irr}$ for strand C and D (15 and 30 nm SiC).

## List of Figures

Figure 1. Resistive transition (vs $B$) at 4.2 K for strand A (200 nm SiC doped $MgB_2$) heated at 700°C and 800°C for various times. $B_{c2}$ is taken at 90% of the normal state response and $\mu_0 H_{irr}$ is taken at 10% of normal state.

Figure 2. Resistive transition (vs $B$) at 4.2 K for strand B (200 nm SiC doped $MgB_2$) heated at 700°C (solid), 800°C (dashed), and 900°C (dot-dash) for various times. $B_{c2}$ is taken at 90% of the normal state response and $\mu_0 H_{irr}$ is taken at 10% of normal state.

Figure 3. Curves of $\mu_0 H_{irr}$ (4.2 K) and $B_{c2}$ (4.2 K) for strand B heated at 700°C, 800°C, and 900°C for various times.

Figure 4. Resistivity vs T at zero field for B-series strands.

Figure 5. Resistive transition (vs $B$) at 4.2 K for strands C and D (15 nm and 30 nm SiC doped $MgB_2$) heated at various times and temperatures. $B_{c2}$ is taken at 90% of the normal state response and $\mu_0 H_{irr}$ is taken at 10% of normal state.

Table 1. $\mu_0H_{irr}$ (4.2 K) and $B_{c2}$ (4.2 K) for strand A (200 nm SiC).

| Sample Name | Tracer ID | HT | $\mu_0H_{irr}$, T | $B_{c2}$, T |
|---|---|---|---|---|
| A-700C/10 | HTR398 | 700°C/10 min | 19.7 | 21.6 |
| A-700C/20 | HTR398 | 700°C/20 min | 19.8 | 22.8 |
| A-700C/30 | HTR398 | 700°C/30 min | 20.3 | 23.9 |
| A-800C/10 | HTR398 | 800°C/10 min | 22.5 | 25.6 |
| A-800C/20 | HTR398 | 800°C/20 min | 21.3 | 24.7 |
| A-800C/30 | HTR398 | 800°C/30 min | 25.4 | 29.7 |

Table 2. $\mu_0H_{irr}$ (4.2 K) and $B_{c2}$ (4.2 K) strands C and D (15 nm and 30 nm SiC).

| Sample Name | Tracer ID | SiC powder size, nm | HT | $\mu_0H_{irr}$, T | $B_{c2}$, T |
|---|---|---|---|---|---|
| D-640C/30 | S1 | 30 | 640°C/30 min | 26.4 | 30.4 |
| C-640C/30 | S2 | 15 | 640°C/30 min | 24.2 | 28.2 |
| D-680C/30 | S3 | 30 | 680°C/30 min | 27.0 | 31.2 |
| D-725C/30 | S5 | 30 | 725°C/30 min | ≈28 | >33 |
| C-725C/30 | S6 | 15 | 725°C/30 min | ≈29 | >33 |

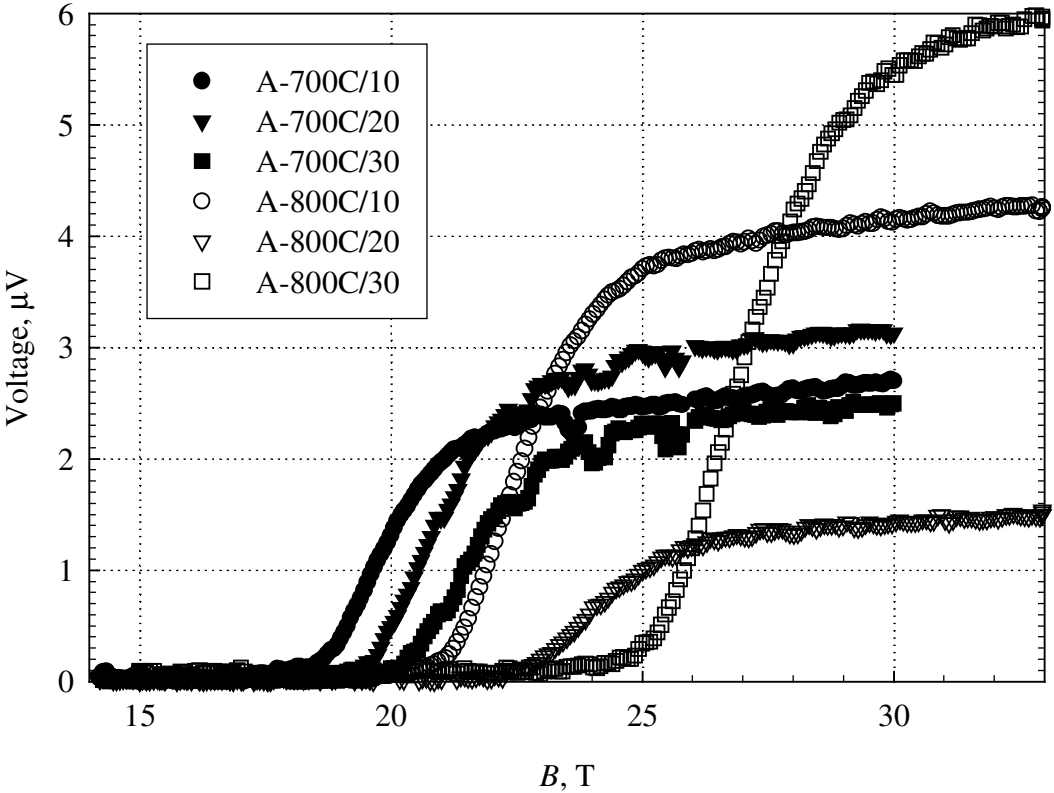

Figure 1.

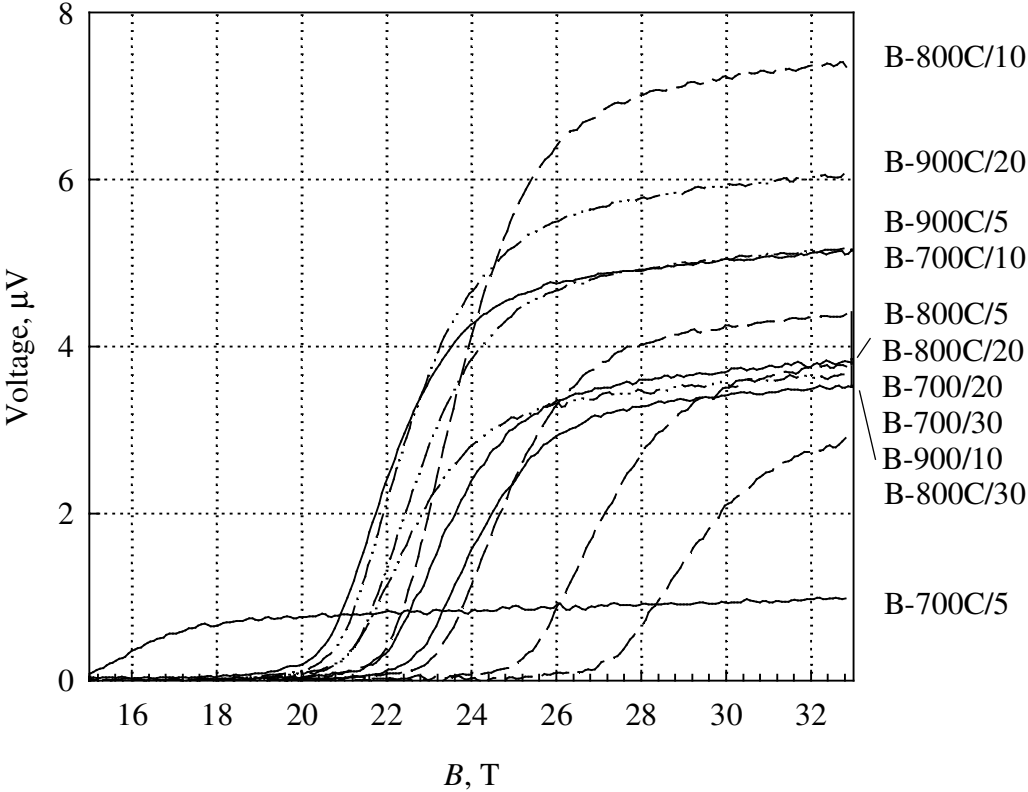

Figure 2.

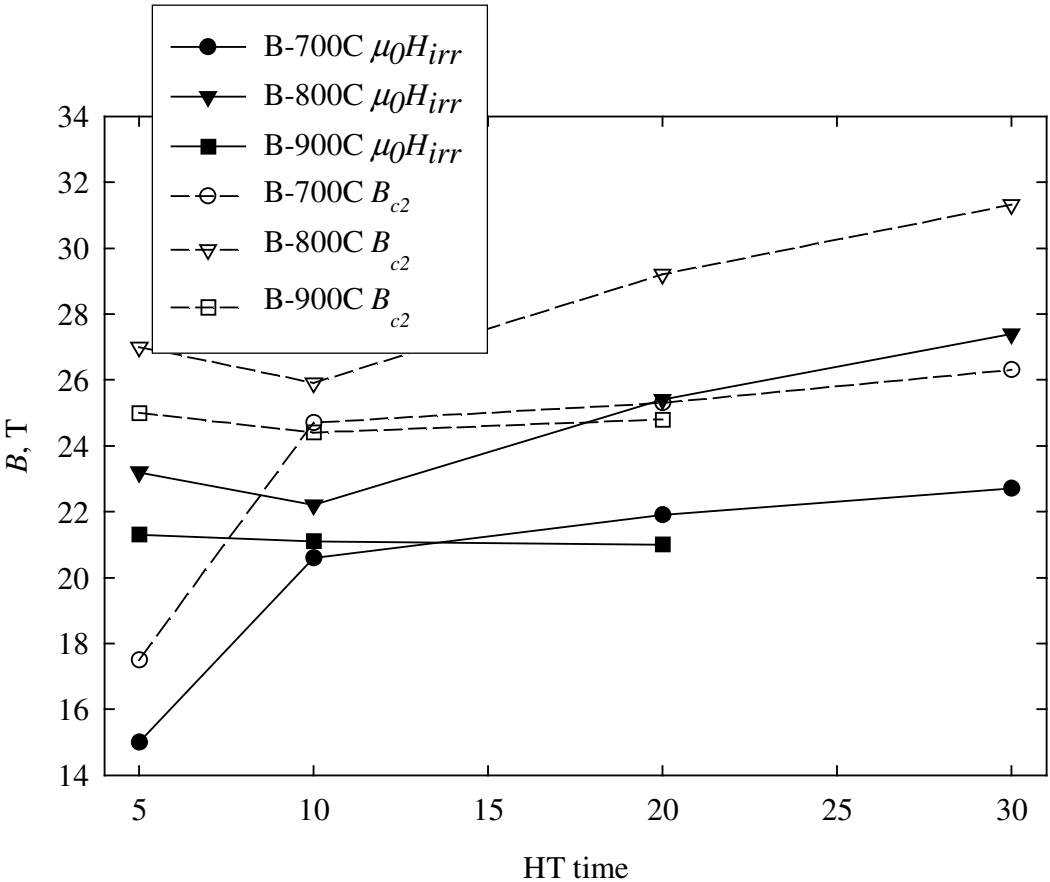

Figure 3.

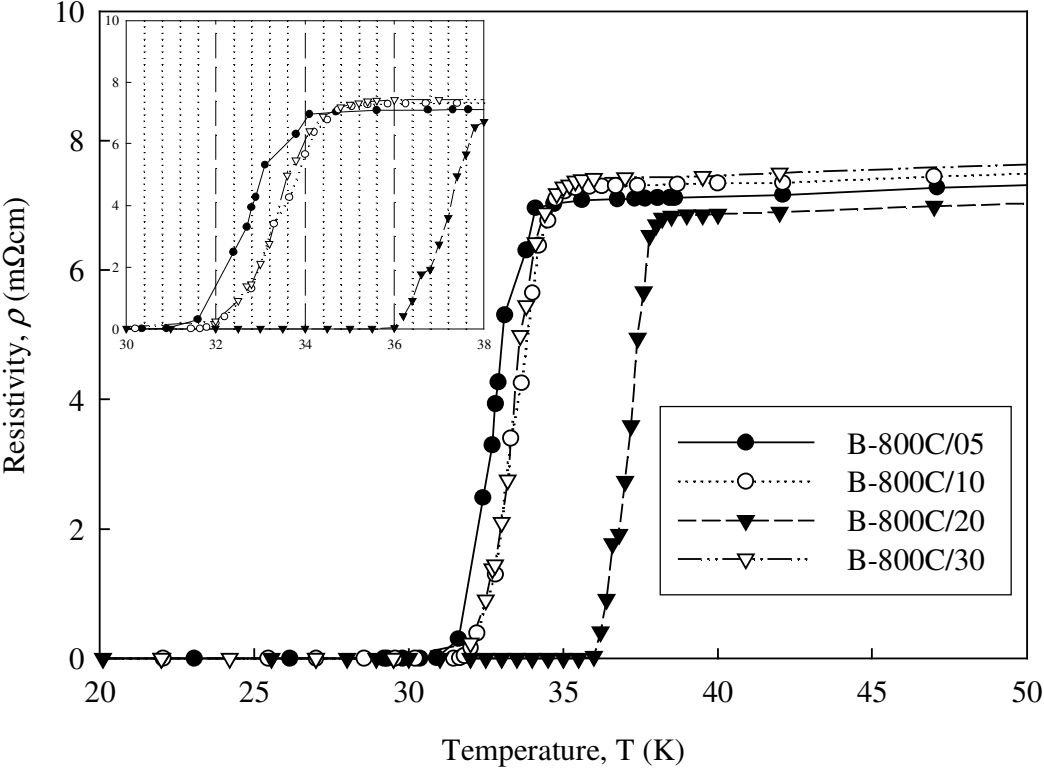

Figure 4.

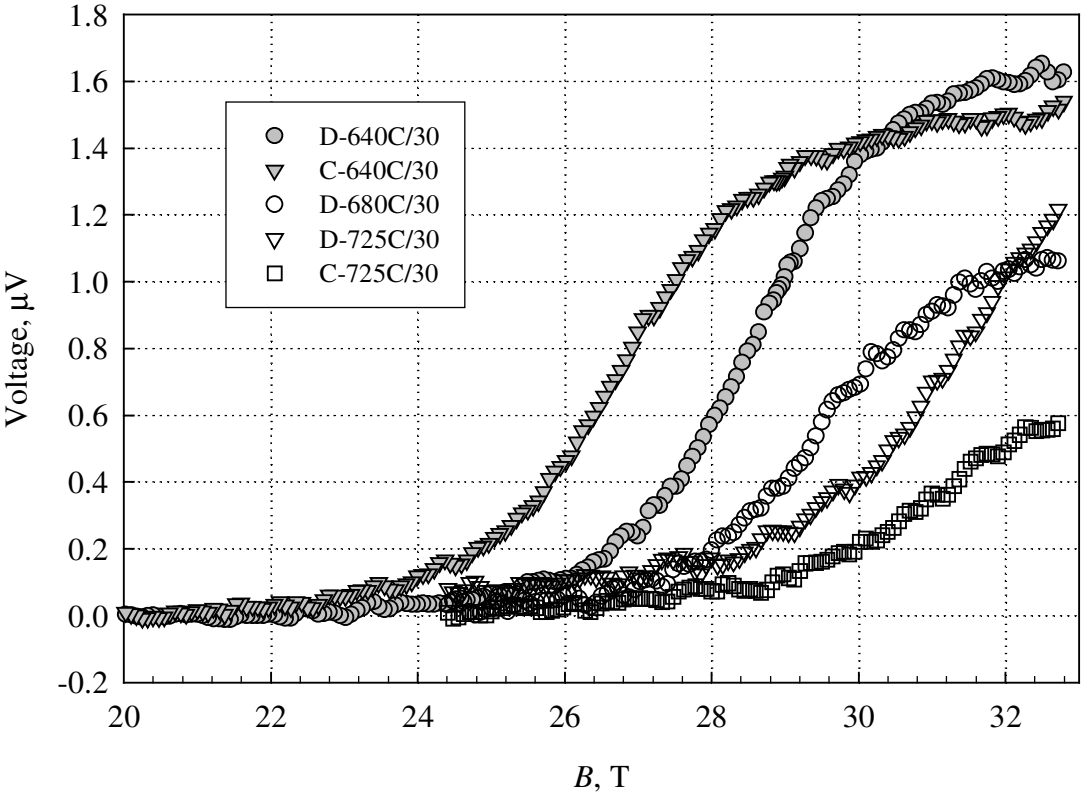

Figure 5.